# Benchmarking system-level performance of passive and active plasmonic components: integrated circuits approach

Alexey V. Krasavin[*] and Anatoly V. Zayats

*Department of Physics, King's College London, Strand, London WC2R 2LS, United Kingdom*
[*]*alexey.krasavin@kcl.ac.uk*

**Abstract:** Using criteria of bandwidth and energy consumption for signal guiding and processing, system-level figures of merit for both passive and active plasmonic circuit components are introduced, benchmarking their performance for the realisation of high-bandwidth optical data communication on a chip. The figure of merit for passive plasmonic interconnects has been derived in terms of the system level performance of the plasmonic circuitry, emphasising the bandwidth and power consumption densities. These parameters are linked to the 'local' waveguide characteristics, such as the mode propagation length, bend radius and mode size. The figure of merit enables a comparison of the main types of plasmonic waveguides and can serve as a benchmark for future designs of photonic integrated circuits. A figure of merit for active photonic- or plasmonic-based electro-optical, thermo-optical and all-optical modulators is also derived to reflect the same benchmarking principles. A particular emphasis is made on establishing a practically oriented benchmark where the integral performance of the circuit, not the size or energy consumption of individual components, plays the defining role.

## 1. Introduction

Microprocessor technology has followed an exponential growth in the computational power for several decades, which is vividly expressed in the empirical Moore's law. The integration level of electronic circuits has already reached the level of ~10 nm. At such dimensions, however, it starts to meet its fundamental limitations. While the operational speed and power consumption of individual MOSFET (metal–oxide–semiconductor field-effect) transistors improve upon miniaturization, the performance of interconnects linking them follows the opposite trend [1,2]. With the size reduction, the drastically increased resistance due to the electron scattering on the interconnect boundaries, accompanied by the increase of the lumped capacitance, result in both higher dissipation losses and longer RC delay times. The latter lowers an achievable data exchange rate, while the former leads to parasitic energy consumption.

A new paradigm for inter- and on-chip data traffic can be based on methods already developed for long-haul optical communications [3-5], using photonic waveguides with their unprecedented bandwidth. The cross-sectional size of traditional optical waveguides is however inherently limited by the diffraction limit of light, which leads to the fundamental mismatch between the integration level of electronic and photonic circuits by almost 3 orders of magnitude. Implementation of surface plasmon polaritons—electromagnetic waves coupled to free electron oscillations localized at a dielectric-conductor interface [6]—as signal carriers may provide a solution for eliminating the mismatch [7]. Additionally, the plasmonic approach takes advantage of the enhanced light-matter interactions in the vicinity of metallic nanostructures where the field is localised and enhanced, offering a way to create much more compact nanoscale active components for switching, modulating and conditioning of optical signals using nanoscale electro-optical and nonlinear optical phenomena [8-10], and furthermore bringing ultra-fast operation speeds [11,12] and ultra-low energy consumption. Plasmonic components, however, introduce significant propagation and insertion losses in photonic circuitry. The important question is, therefore, whether the advantages in the circuit miniaturization, speed increase and operating energy consumption reduction provided by plasmonic components outweigh the penalty of the energy dissipation in individual plasmonic active and passive components when considering the photonic system as a whole.

In this article, we address this question by deriving, from the bandwidth and energy consumption considerations of the integrated photonic system, benchmarking parameters (figures of merit, FOMs) characterising the performance of a) plasmonic waveguides for the on-chip data communication and b) active plasmonic components, particularly, modulators. Importantly, for the passive components the figure of merit was obtained on the basis of most general and practically relevant considerations of the data traffic rates and power dissipation densities, which appeared to be related to the broadly used local waveguide characteristics, such as the bandwidth, signal propagation length and integration parameters. This allowed the generalisation of all earlier FOM considerations, and reveals a previously overlooked bandwidth factor. On this basis, the performance of the main types of plasmonic waveguides was benchmarked. In the second part of the paper, a figure of merit for optical modulators has been derived. Taking advantage of the localisation of optical signals at nanoscale dimensions, the plasmonic electro-optical and all-optical modulators have extremely small sizes, which leads to their extremely high operational speeds (possibly tens of Tb/s) and low energy consumption (down to few fJ per bit). The FOM of plasmonic modulators reveals their qualitative advantages compared to the best traditional state-of-the-art photonic counterparts.

## 2. Figure of merit for passive plasmonic waveguides

Considering the advantages offered by the plasmonic approach in the size reduction of nanophotonic components, a broad variety of plasmonic waveguiding geometries has been proposed, including long-range [13], dielectric-loaded [14], hybrid [15,16], nanowire [17], metal-insulator-metal (MIM) and wire-MIM [8] plasmonic waveguides, to name just a few (Fig. 1). These designs cover the whole range of mode sizes achievable in the optical communication wavelength range from micrometers, as in long-range plasmonic waveguides (Fig. 1(a)), to tens of nanometres as, e.g., in wire and wire-MIM plasmonic waveguides (Fig. 1 (d,e)). For all these waveguides, however, the achieved signal localisation is accompanied by intrinsic Ohmic losses introduced by the metallic elements of the waveguides, resulting in a propagation loss of the plasmonic signal. The propagation length of the signal, therefore, ranges from a centimetre scale (as in long-range plasmonic waveguides) for the waveguides with lower signal localisation to a micrometer scale for the waveguides with the highest confinement, e.g., in wire and wire-MIM plasmonic waveguides. In fact, there is a general trade-off between these two characteristics: higher mode localisation usually corresponds to a larger relative part of the mode energy propagating in the metal, which leads to higher losses and, therefore, a lower propagation length.

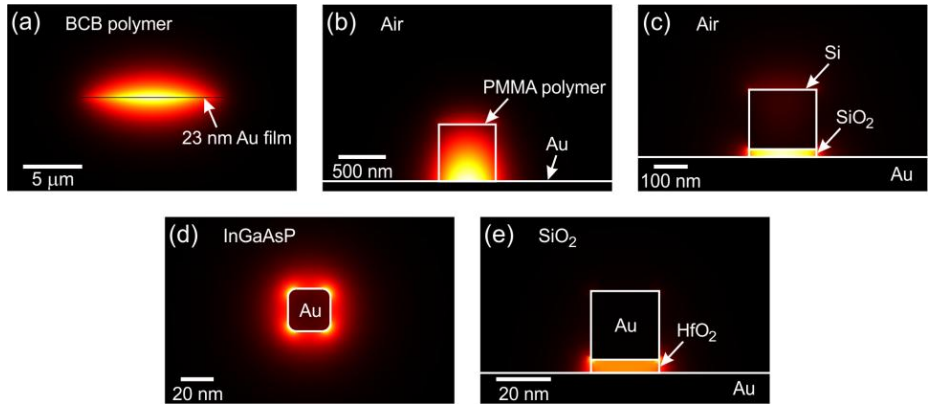

**Figure 1.** Electric field profiles $|\mathbf{E}|^2$ for the modes in (a) long-range, (b) dielectric-loaded, (c) hybrid, (d) nanowire (asymmetric mode), and (e) wire-MIM plasmonic waveguides. For easy comparison, the results are presented for the same plasmonic platform (Au) and operational wavelength ($\lambda = 1550$ nm). The choice of embedding dielectric (indicated in the figure) is typical for each type of the waveguide.

In order to compare the diverse variety of plasmonic waveguides, a figure of merit can be introduced to quantitatively characterise their guiding performance. The first FOM to characterize the performance of various passive waveguides was introduced by Buckley and Berini [18]. As the best starting point, the figure of merit

$$M_1 = 2\sqrt{\pi} \cdot \frac{L_{prop}}{\sqrt{S}} \qquad (1)$$

represented a straightforward measure of the above-mentioned trade-off: a ratio between the signal propagation length $L_{prop}$ (after which the mode intensity decreases $e$ times), reflecting the propagation characteristic of the mode and the size of the mode $\sqrt{S}$ (with $S$ being the effective mode area), reflecting the mode confinement. Although it has a clear physical meaning and gives a vivid characterization of the mode as such, this FOM is not ideal for benchmarking the performance of the waveguides implemented in highly-integrated optical circuits. The square root of the mode area in the denominator of $M_1$ is effectively a measure of how closely parallel waveguides can be placed on a chip. However, modes with the same effective areas can couple with different efficiency to a neighbouring waveguide depending on their spatial field distributions. Furthermore, it was shown that there is a variety of ways to quantify the effective mode area, frequently giving essentially different and even opposite results (for an extensive overview see Ref. [19]).

In the further development, the figure of merit was modified in order to include a direct measure of the coupling efficiency [17,20]. When two parallel waveguides are placed next to each other, their coupling leads to a gradual transfer of the mode from one waveguide to another with full energy relocation after a distance $L_{coupl}$. Taking this into account, the waveguide centre-to-centre separation $d_{sep}$, representing how close the two identical waveguides can be placed, so that the coupling distance $L_{coupl}(d_{sep}) = 4L_{prop}$, will give a direct measure of the achievable integration density: after one propagation length, only 15% of the mode energy will be coupled to the neighbouring 'aggressor' waveguide. Such a condition, expressing the coupling length in terms of the propagation length, was chosen to compare diverse guiding approaches, embracing an extremely wide range in the propagation-confinement parameter space. (Here we note that although this definition is universal, it is postulated, leaving the degree of freedom for adjusting it to a particular application, for example, instead of $L_{prop}$ a characteristic interconnect length $L_{int}$ can be used if $L_{prop} > L_{int}$.) Using the coupling distance considerations, the figure of merit can be represented as [20]:

$$M_{loc}^{lin} = \frac{L_{prop}}{d_{sep}}. \qquad (2)$$

Notably, although the coupling phenomenon involves mode evolution in the third dimension along the waveguides, seemingly requiring 3D numerical simulations, the coupling characteristic ($L_{coupl}(d_{sep})$) can be derived from a 2D eigenmode analysis of two coupled waveguides [20]. Particularly, this can be done by monitoring the difference between mode effective indexes of the symmetric ($n_{eff}^{sym}$) and antisymmetric ($n_{eff}^{asym}$) modes appearing in such a system: $L_{coupl} = \lambda / \left[ 2\left(n_{eff}^{sym} - n_{eff}^{asym}\right) \right]$, where $\lambda$ is the operational wavelength. Therefore, it has the same numerical simulation complexity as the usual modal area estimation.

The second main parameter characterising particularly the performance of a multi-branched waveguide circuitry is an optimal bend radius, minimising signal loss along a curved waveguide section. This parameter defines the size of all circuit components, such as splitters, waveguide ring resonators (WRRs), Mach-Zehnder interferometers (MZIs), etc. The optimal waveguide radius is a trade-off between Ohmic loss (higher for bigger radii) and radiation losses (higher for smaller radii) [21,22]. A figure of merit $M_2 = 2L_{prop} \cdot (n_{eff} - n_{surr})$ [18], where $n_{eff}$ is the mode effective index and $n_{surr}$ is the refractive index of the surrounding medium, gives only a partial answer, introducing an approximate measure of only the radiation losses: the further the SPP mode dispersion is from the light line, the lower is the coupling efficiency to the escaping light. At the same time, the radiation losses are also influenced by the particular waveguide geometry, and can be different for waveguides with the same value of the FOM $M_2$. As an alternative approach, the figure of merit for multi-branched plasmonic circuitry was proposed in Ref. [17], taking into account both Ohmic and radiation losses and directly estimating the waveguide bends performance:

$$M_{loc} = \frac{L_{prop}}{d_{sep}} \cdot \left(\frac{T(r)}{r}\right)_{max}, \qquad (3)$$

where $\left(T(r)/r\right)_{max}$ is the maximum value of the ratio of the transmission $T(r)$ and bend radius $r$, which can be found varying the radius in 3D numerical simulations.

The above figures of merit were derived using 'local' characteristics of the waveguide performance: the propagation length, cross-talk between neighbouring waveguides and bent waveguide performance. The most relevant to practical applications, however, is the estimation of the system-level performance of the multi-branched plasmonic circuit, when considered as a complete data communication architecture. The most general benchmark describing the performance from this point of view can be defined considering the bandwidth and energy requirements for transferring a signal:

$$M_{syst} = \frac{b}{p}, \qquad (4)$$

where $b$ is the bandwidth density which is a number of bits transferred per second per unit area of an integrated circuit [23] and $p$ is the power density required to sustain the bit-rate (in the case of plasmonic waveguides this is the power lost in transmission due both radiation and absorption losses). To estimate such a figure of merit related to system-level performance, we consider two extreme scenarios for the network layout (Fig. 2).

In the first case, the circuit consists solely of straight waveguides, connecting input and output nodes. We will consider the most dense layout where the distance between the nodes is the smallest and equal to the separation distance between two parallel waveguides $d_{sep}$ (Fig. 2(a)). Using the smallest unit cell $2d_{sep} \times 2d_{sep}$, the highest bandwidth density with straight interconnects achievable for each type of the waveguides can be estimated: $b = B/d_{sep}^2$, where $B$ is the highest bandwidth (i.e. the highest bitrate which can be transmitted for a waveguide of a given length). It should be noted that a bitrate and, therefore, a bandwidth density can be further increased using, e.g., wavelength division multiplexing (WDM) via transmitting several bit streams through the same waveguide simultaneously at different carrier wavelengths, or other techniques. The total bandwidth is then $B = NB_1$, where $N$ is the number of channels, considered to be equal for all plasmonic waveguides and $B_1$ is the bandwidth of a single data stream. The bandwidth $B_1$ is limited for a single-mode waveguide by the group velocity dispersion (GVD) and nonlinear effects which differently affect the propagation conditions for different spectral components of the signal, leading to its distortion. In lossy waveguides, broadening of the pulses depends on the second-order dispersion of both real and imaginary parts of the propagation constant $\beta$ [24]:

$$B_1 \sim L^{-1/2} \left( \left( \frac{\partial^2 \operatorname{Re}\{\beta\}}{\partial \omega^2} \right)^2 + \left( \frac{\partial^2 \operatorname{Im}\{\beta\}}{\partial \omega^2} \right)^2 \right)^{-1/4}. \qquad (5)$$

For the plasmonic waveguides compared below, it was found that $\left(\partial^2 \operatorname{Im}\{\beta\}/\partial \omega^2\right)^2 \ll \left(\partial^2 \operatorname{Re}\{\beta\}/\partial \omega^2\right)^2$, which is determined by both the material and modal dispersions. Since $\partial^2 \operatorname{Im}\{\beta\}/\partial \omega^2$ is 1 or 2 orders of magnitude smaller than $\partial^2 \operatorname{Re}\{\beta\}/\partial \omega^2$, depending on the waveguide type, and leads to the correction to the bandwidth less than 0.3%, it can be neglected, and the bandwidth can be estimated as [25]:

$$B_1 = \left( 8\pi L \frac{\partial^2 \operatorname{Re}\{\beta\}}{\partial \omega^2} \right)^{-1/2} = \left( 8\pi \frac{L}{\upsilon_g^2} \frac{\partial \upsilon_g}{\partial \omega} \right)^{-1/2}, \qquad (6)$$

where $\upsilon_g$ is the group velocity of the mode and $L$ is the interconnect length. For the estimation of the bandwidth, the length of the plasmonic interconnect was taken to be equal to the propagation length $L = L_{prop}$. From the discussion related to Eqs. (5) and (6) it follows, that for the description of pulse propagation in the considered cases it is possible to apply the concept of the group velocity, otherwise the approach based on the energy velocity should be implemented. The nonlinear effects (affecting the bandwidth in the case of optical fibres) were not taken into account because for much smaller interconnect lengths and lower targeted pulse energy of approximately 1 fJ these effects are small even considering the larger nonlinearity in gold and the field enhancement in the plasmonic mode [26,27].

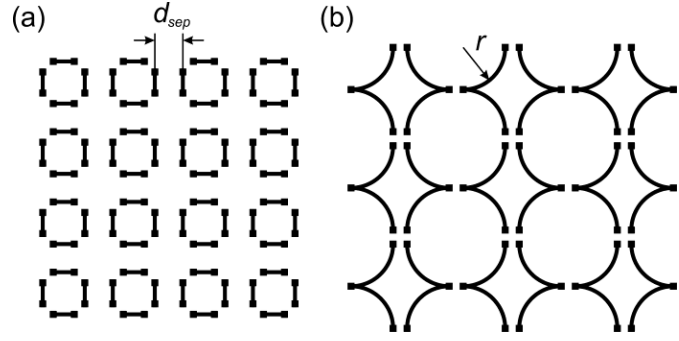

**Figure 2.** Two extreme scenarios for data network layout: (a) the network consists of only straight waveguides and (b) the network consists of only circularly curved waveguides relevant to waveguide bends, splitters and ring-resonators (in the case of waveguide bends and splitters, one can get approximately 10% improvement in performance using harmonic function based shapes [14,22]). The minimum curvature radius $r$ with still affordable radiation losses is usually bigger than the separation distance $d_{sep}$ with the affordable cross-talk.

Considering the same average power $p_0$ sent through each interconnect, the power loss per unit area can be estimated as

$$p = p_0\left(1-\exp\left[-d_{sep}/L_{prop}\right]\right)/d_{sep}^2 \approx p_0/(d_{sep}L_{prop}), \tag{7}$$

where the fact that $d_{sep} \ll L_{prop}$ was taken into account. This results in the following system-level figure of merit:

$$M_{syst}^{lin} = \frac{b}{p} \sim B_1 \frac{L_{prop}}{d_{sep}}. \tag{8}$$

In the second case (Fig. 2 (b)), if the circuitry is predominantly multi-branched, both the bandwidth density and the power loss will be primarily defined by the performance of the waveguide bend. Here, as a universal building shape for waveguide bends, splitters and ring-resonators, a bend shape is considered in a form of a circular arc. In this case, the bandwidth per unit area is equal to $b = B/r^2$, while the corresponding loss density is equal to $p = \left(1-\exp\left[-1/4 \cdot 2\pi r/L_{prop}^{bend}(r)\right]\right)/r^2$, where $r$ is the bend radius and $L_{prop}^{bend}(r)$ is the propagation length of the mode along the bend, determined by both Ohmic and radiative losses. Taking into account that the transmission through a bend is usually reasonably high (~ 0.75-0.95), and, thus, $-1/4 \cdot 2\pi r/L_{prop}^{bend}(r) \ll 1$, we can approximate $p \approx 1/2 \cdot \pi L_{prop}^{bend}(r)/r$. The figure of merit for such circuit is then given by

$$M_{syst}^{circ} = \frac{b}{p} \sim B_1 \left(\frac{L_{prop}^{bend}(r)}{r}\right)_{max}, \tag{9}$$

where $\left(L_{prop}^{bend}(r)/r\right)_{max}$ is the optimal (highest) value of the ratio of the propagation length along the bend and its radius for a given waveguide type. Finally, in the case of an arbitrary circuitry layout, we can combine these two figures of merit and obtain

$$M_{syst} = B_1 \cdot \frac{L_{prop}}{d_{sep}} \cdot \left(\frac{L_{prop}^{bend}(r)}{r}\right)_{max}. \tag{10}$$

Strikingly, although the local (Eq. (3)) and the system-level figures of merit are derived from absolutely different perspectives, they reflect absolutely the same dependences on the waveguide performance characteristics. The only difference is that in the former case the bend transmission-to-radius ratio $T(r)/r$ is maximized, while in the latter case, the same should be done for the propagation characteristic of the bend section $L_{prop}^{bend}(r)/r$. It should be noted that the transmission through the bend $T(r)$ and the propagation length along the bend $L_{prop}^{bend}(r)$ are closely related characteristics. Another important parameter in Eq. (10), which was completely disregarded, is the waveguide bandwidth, which is revealed using the system-level performance approach. Due to small interconnect lengths, the calculated single-stream bandwidth $B_1$ reaches extremely high values of tens of Tb/s (Table 1), and practically will be limited by available data modulation and detection electronic technologies presently capable of 10—100 Gb/s modulation rates.

The derived system-level figure of merit was evaluated to compare the performance of the main types of plasmonic waveguides (Table 1). The relevant waveguide characteristics were found using 2D eigenmode and full 3D finite element numerical simulations. For fair comparison, all the waveguides were implemented on a gold material platform at telecommunication wavelength $\lambda = 1550$ nm. Additionally, for the wire-MIM waveguide an Al platform was used to compare the performance of the original design [8] and for the wire waveguides, two different dielectric coatings were implemented. The optical constants of metals were taken from Ref. [28]. As expected (Table 1), for all the waveguides a trade-off between $L_{prop}$ and $d_{sep}$ is observed. The propagation length along the ideally curved section $L_{tot}$ is a factor of 1.2—4 shorter than $L_{prop}$ as it also includes the radiation losses. The ideal radius can be of the same order of magnitude as the waveguide cross-sections, as in the case of the most highly-integrated waveguides, or can be up to 3 orders of magnitude larger, as in the case of the long-range SPP waveguides. There is a clear trend of the increase of the waveguide bandwidth for the waveguides allowing a higher integration level, which can be partly explained by a shorter distance along which the pulse needs to propagate ($L_{prop}$) and thus GVD has less influence.

An interesting universal tendency encompassing the characteristics of plasmonic waveguides of completely different designs can be observed. For instance: the cross-talk limited photonic integration density of straight waveguides can be different by 2 orders of magnitude, the bend radii can be different by 5 orders of magnitude, propagation lengths can be different by 4 orders of magnitude (cf. $d_{sep}$, $r$ and other parameters for long-range and wire-MIM SPP waveguides), but the system-level figures of merit always fall within 2 orders of magnitude interval from each other. As follows from Table 1, for particular parameters of plasmonic waveguide designs taken from the literature, the hybrid plasmonic waveguides have the highest FOM values; however, one needs to bear in mind that the figure of merit for a given waveguide design may vary with its geometrical parameters (e.g., for hybrid waveguides, the figure of merit can vary by an order of magnitude for different parameters). In this regard, apart from benchmarking the performance of waveguides of various designs, the derived FOM can be used to optimise the performance of a chosen type of the waveguide. For complete assessment, the FOM should be used in conjunction with other technological considerations, such as compatibility of the waveguides' material platform with the existing fabrication processes or the cost of mass production of such circuits, and, in particular, with the specific application requirements, which may prioritise one or another performance characteristic of a waveguide interconnect. (A typical example of these may be given considering, e.g., conventional silicon-on-insulator waveguides with a very low propagation loss which would provide very high figure of merit (Table 1) but their applications are limited to low integration densities of global optical interconnects.)

It also needs to be noted that the insertion losses of nanophotonic circuitries in, e.g., fibre-based networks play an important role for determining the power efficiency of the chosen photonic integrated circuit architecture, but they are similar for all the waveguides considered. Moreover, for fully integrated applications, on-chip nanoscale light sources for generating light and/or SPPs directly into the plasmonic waveguided modes under either electrical [29,30], or optical [31] excitation will be required.

| | Long-range waveguide [13][*] | DLSPPW [14] | Hybrid waveguide | Wire waveguide (asym. mode) [17] | Wire-MIM waveguide [8] | 250 nm × 400 nm SOI waveguide |
|---|---|---|---|---|---|---|
| $B_1$ | 14 Tb/s | 39 Tb/s | 70 Tb/s[**] ([15])<br>20 Tb/s ([16]) | 61 Tb/s (InGaAs coat.)<br>90 Tb/s (SiO$_2$ coat.) | 139 Tb/s (Al)<br>86 Tb/s (Au) | 3.2 Tb/s |
| $L_{prop}$ | 1.75 mm | 45 μm | 45 μm ([15])<br>16.3 μm ([16]) | 310 nm (InGaAs coat.)<br>1.2 μm (SiO$_2$ coat.) | 420 nm (Al)<br>480 nm (Au) | ~1 cm (limited by waveguide wall roughness) |
| $d_{sep}$ | 12 μm | 1.8 μm | 1.05 μm ([15])<br>1.3 μm ([16]) | 85 nm (InGaAs coat.)<br>200 nm (SiO$_2$ coat.) | 65 nm (Al)<br>60 nm (Au) | 1.2 μm |
| $L_{prop}^{bend}$ | 1.1 mm | 31.5 μm | 38 μm ([15])<br>15.1 μm ([16]) | 110 nm (InGaAs coat.)<br>123 nm (SiO$_2$ coat.) | 196 nm (Al)<br>405 nm (Au) | 350 μm |
| $r$ | 3 mm | 6 μm | 2 μm ([15])<br>3.5 μm ([16]) | 12.5 nm (InGaAs coat.)<br>12.5 nm (SiO$_2$ coat.) | 12.5 nm (Al)<br>12.5 nm (Au) | 2.9 μm |
| $M_{syst}$ | 0.75·10$^3$ Tb/s | 6.1·10$^3$ Tb/s | 56.9·10$^3$ Tb/s ([15])<br>1.1·10$^3$ Tb/s ([16]) | 2·10$^3$ Tb/s (InGaAs coat.)<br>5.3·10$^3$ Tb/s (SiO$_2$ coat.) | 14.1·10$^3$ Tb/s (Al)<br>22.3·10$^3$ Tb/s (Au) | 3.2·10$^6$ Tb/s |

**Table 1.** Benchmarking characteristics for various plasmonic waveguides presented in Fig. 1 and comparison of the performance of the waveguides using the system-level figure of merit Eq. (7). [*]Although assumption $-1/4 \cdot 2\pi r / L_{prop}^{bend}(r) \ll 1$ is not valid for long-range SPP waveguides, the calculated figure of merit can be treated as an upper estimate. [**]In all the considered cases, the variation of the group velocity dispersion over the waveguide bandwidth is below 5%, except for the hybrid waveguide from Ref. [15] with about 10% variation due to the higher-order effects (this would affect a FOM by not more than ~10%).

## 3. Figure of merit for active plasmonic components

The advantages which the plasmonic approach brings to the development of electro-optical, thermo-optical or all-optical modulators and switches are related to their small size and increased light-matter interaction strength near plasmonic interfaces [8,32-35], thus reducing the required operational energy and increasing speed. To achieve quantitate comparison of various optical modulators, an active figure of merit is needed which takes into account all application requirements.

The performance of optical modulators is mainly characterised by two parameters: the achievable modulation speed (or modulation bandwidth) and the energy consumption per bit [32]. In the case of electro-optical modulation, the modulation bandwidth is defined by the $RC$-delay of the component:

$$B_{mod} = \frac{1}{\tau} = \frac{1}{RC}, \tag{11}$$

where $R$ is the external resistive load, taken to be the same for all the examples considered below, and $C$ is the capacitance of the device (determined by the electrical system through which the control voltage is applied to the device), which is directly proportional to the characteristic device size $a$. The simplest example is a parallel capacitor, for which the area $S \sim a^2$ and the gap between plates $d \sim a$, and therefore the capacitance $C = \varepsilon_0 \varepsilon S/d \sim a$ (here, $\varepsilon_0$ and $\varepsilon$ are permittivities of vacuum and the material between plates, respectively). This explicitly shows the advantage of nanoscale sizes of plasmonic modulators for the increase of the modulation speed.

The energy consumption per bit can be estimated as the energy required to charge the effective device capacitor to the required voltage $V_{3dB}$ returning 3 dB intensity modulation through the induced change of the mode phase or absorption [36]

$$P = \frac{1}{4} C V_{3dB}^2. \tag{12}$$

Since the control voltages of the plasmonic modulators are of same order of magnitude as for their conventional photonic counterparts [8, 32-35,37,38], a requirement $P \sim C \sim a$ underlines a crucial advantage offered by the plasmonic approach: reducing the modulator size leads to the reduced energy consumption.

In addition to the above main characteristics, other important operational parameters should be considered such as contrast ratio, optical bandwidth (the wavelength range in which the modulator can operate, which also affects the modulation speed), insertion loss, footprint and the optical power it can accommodate. As a consequence, a huge variety of figures of merit benchmarking the performance of optical modulators exist proposing different characteristics as the most important as well as relevant to different modulator designs. At the same time, a universal modulator FOM capable of comparing various designs and approaches from a point of view of system-level performance is absent.

A good starting point for the development of such a figure of merit is the most common benchmark, used for conventional phase-shift electro-optical modulators based on Mach-Zehnder interferometers [39]:

$$M_1^{act} = \frac{1}{V_{3dB} l}, \tag{13}$$

where $l$ is the modulator length. This FOM can also be applied to absorption-based modulators, while a figure of merit of the same form $M_1^{act} = 1/(V_{3dB} r)$ can be introduced for another common phase-shift design based on a ring-resonator, where $r$ is the ring radius characterising the device optical path. An advantage of $M_1^{act}$ is that it brings to the focus the efficiency of modulation *design* and the performance of the *materials* rather than the overall device performance. It however completely disregards such an important parameter as the switching speed (modulation bandwidth $\Delta f$). As an alternative, another figure of merit has been proposed [40]:

$$M_2^{act} = \frac{B_{mod}}{V_{3dB}}. \tag{14}$$

Using Eq. (11), one can rewrite it as

$$M_2^{act} = \frac{1}{V_{3dB}RC} \sim \frac{1}{V_{3dB}Rl} \sim M_1^{act} \tag{15}$$

and conclude that for the case of a fixed waveguide cross-section it is analogous to $M_1^{act}$.

The difficulties arise when one tries to apply these figures of merit to assess the practical performance of the component when integrated in circuitry and uses the energy consumption as the modulator characteristic. For example, between the two modulators of the same cross-section having the same values of $M_1^{act}$, the one which has twice smaller $l$ (and therefore twice larger $V_{3dB}$) will have a power consumption of two times higher, as it follows from Eq. (12). Therefore, considering the practical performance of the modulator as an integral part of a chip, one logically comes to the widely used FOM based on the electrical power required for component switching, rather than voltage [41]:

$$M_3^{act} = \frac{B_{mod}}{P}. \tag{16}$$

Finally, including in the figure of merit a factor which became particularly important with the development of the plasmonic-based modulators — modulator on-state attenuation due to the increased optical losses [42], expressed through transmission coefficient in the on-state $A$ — we arrive to the final expression for the modulator figure of merit on the basis of the system-level performance:

$$M_4^{act} = A\frac{B_{mod}}{P}. \tag{17}$$

Using Eqs. (11) and (12), it can be simplified to $M_4^{act} = 4A\left(RV_{3dB}^2 C^2\right)^{-1} \sim a^{-2}$. This immediately confirms the advantages of plasmonic electro-optic modulators, having the potential for extremely small sizes $a$ and is further illustrated by benchmarking various designs (Table 2). It is also worth to note that due to the universal character of the parameters in $M_4^{act}$, it can also be applied for other types of optical modulators, e.g. all-optical, thermo-optical or opto-mechanical.

If one divides both numerator and denominator of Eq. (17) by the modulator footprint $S$,

$$M_4^{act} = A\frac{B_{mod}/S}{P/S} = A\frac{b_{mod}}{p}, \tag{18}$$

one can see another meaning of the figure of merit: it presents a balance between the active bandwidth density $b_{mod}$ and the power density $p$ required to achieve it. In other words, the figure of merit remains the same if the increased integration density of the switching components is balanced by a proportional increase in switching energy consumption of a circuit as a whole. If, however, the integration level is a major consideration so that the components with smaller footprint become preferable, the higher rate of increase of the circuit energy consumption should be allowed for.

|  | Commercial LiNbO$_3$ modulator | Si ring resonator modulator [37] | GeSi electro-absorption modulator [38] | Electro-optical plasmonic modulator [33] | Field-effect plasmonic modulator | Plasmonic all-optical modulator [9] | Si-photonics nonlinear demultiplexer [43] | Metamaterial-based all-optical modulator [44] |
|---|---|---|---|---|---|---|---|---|
| $B_{mod}$ | 40 Gb/s | 10 Gb/s | 40 Gb/s | 400 Gb/s | 250 Gb/s (10 Tb/s) ([34]) 15 Tb/s ([8]) | 15 Gb/s | 1 Tb/s | 0.1—1 Tb/s |
| $P$ | 10 pJ | 50 fJ | 60 fJ | 60 fJ | 4 fJ (30 aJ) ([34]) 15 aJ ([8]) | 0.3 fJ | 1 pJ | 3.7 pJ |
| $S$ | 5 cm$^2$ | 1000 μm$^2$ | 500 μm$^2$ | 30 μm$^2$ | 1 μm$^2$ ([34]) 10$^{-2}$ μm$^2$ ([8]) | 300 μm$^2$ | 1 μm × 4 mm | 0.5 μm$^2$ |
| $A$ | 0.3[*] | 0.6 | 0.3 | 0.1 | 0.8 ([34]) 0.6 ([8]) | 0.1 | 1 | 0.3 |
| $B_{opt}$ | 40 nm | 0.1 nm | 35 nm | >100 nm | ≫100 nm | ≫100 nm | n/a | 20 nm |
| $M_4^{act}$ | ~0.001 Gb/s/fJ | 0.12 Gb/s/fJ | 0.2 Gb/s/fJ | 0.7 Gb/s/fJ | 50 Gb/s/fJ (3·10$^5$ Gb/s/fJ) ([36]) 6·10$^5$ Gb/s/fJ ([8]) | 5 Gb/s/fJ | 1 Gb/s/fJ | 0.01-0.1 Gb/s/fJ |

**Table 2.** Benchmarking characteristics for various modulators and comparison of the performance of the modulators using figure of merit in Eq. (17). Table values for the modulator from Ref. [34]: without brackets – conservative estimation from the original article, in brackets – potential value for an optimised design. [*] The value of $A = 0.3$ corresponds to 30% modulator transmission in the on-state.

The impact of the optical bandwidth $B_{opt}$ on the performance of the optical modulator is more difficult to account for, as it has a technological character (fabrication precision, temperature stability, etc.), therefore, this parameter was not included in the figure of merit. Apart from this, in highly resonant systems, such ring-resonators or photonic crystal cavities, the low optical bandwidth (corresponding to a long resonance build-up time) decreases the modulation bandwidth $B_{mod}$, which competes with the modulation bandwidth defined by the *RC*-delay, and can become the major factor limiting the overall bandwidth value.

Several observations can be made from the data in Table 2. Two mainstream designs of electro-optic modulators (columns 3 and 4) which are under consideration to replace the commercial LiNbO$_3$ modulators (column 2), namely Si ring resonator modulators, based on a free carrier dispersion effect, and GeSi modulators based on an electro-absorption effect do offer better performance in the system-level settings. These three types of modulators have comparable modulation bandwidths and most of the other parameters, but smaller size and, consequently capacitance, of the Si and GeSi based designs ensures lower energy consumption characteristics, which are on fJ rather than pJ scale. Such a performance improvement is reflected in a straightforward way in the hundred-fold increase of the figure of merit (here, one needs to consider that the optical bandwidth of the Si-based modulator is rather narrow due to the use of a highly resonant system). Another several times increase in the figure of merit can be achieved implementing the plasmonic approach, which provides further reduction of the device size, combining it with an efficient active material platform (a polymer with electro-optical Pockels effect). The main role in the FOM improvement is played here by an order of magnitude higher modulation bandwidth, while the energy consumption remained at the same level due to a rather high (5—7.5 V) modulation voltage and additional price for the plasmonic design is paid in a form of absorption losses (cf. column 5 and columns 3 and 4). In the examples of field-effect plasmonic modulators (column 6), the nanoscale dimensions of the device and, particularly, the extreme localisation of plasmonic modes was combined with extremely strong electro-absorption effect, sufficient to act on the modes at the nanoscale. Thus, the obtained very-low-capacitance devices provide both high (Tb/s range) modulation bandwidth, fJ (potentially aJ) power consumption per bit; all these parameters result in the highest figure of merit among the considered electro-optical modulators. To demonstrate the universal character of the derived figure of merit, several examples of all-optical modulators were considered (columns 7—9). Here, due to variety of available designs and material platforms, the operational characteristics span wide range of parameters: modulation bandwidth varies from Gb/s to Tb/s, power consumption from sub-fJ to a few pJ, with up to 2 orders of magnitude difference in sizes. Generally, the presented figures of merit are comparable to that of the considered electro-optic modulators, but in our view they should be benchmarked within their own all-optical domain, since they present a completely different all-optical chip paradigm than electro-optical devices. In the future, all-optical modulators have a potential of achieving the highest figures of merit due to very high modulation rates available with third-order nonlinear processes [43,45,46].

Finally, to complete the overview of available and future switching and modulation technologies, thermo-optical and nano-opto-mechanical approaches should be mentioned. Thermo-optical switches are low-speed (KHz—MHz range) devices in both photonic [46] and plasmonic [47] realizations, important for some applications in signal routing and variable-optical attenuators. Potentially, plasmonic-based devices should provide better FOMs than dielectric photonic ones due to smaller sizes and, thus, reduced power consumption. However, their fair comparison was not possible with the available published data due to very different thermo-optical materials used. Nano-mechanical approaches for photonic modulators and switches may provide some of the lowest switching energies per bit with moderate MHz-range speeds in both photonic and plasmonic environments [48-51]. Potential use of new plasmonic materials in nanophotonic integrated circuits [52] may further benefit their size and energy reduction, and thus bring improvements of the system-level FOM.

## 4. Conclusion

Figures of merit for passive plasmonic interconnects and active components (modulators) were derived on the basis of system design requirements and the most general considerations describing their efficiency to realise high-bandwidth data communication in multi-branch photonic circuitry. Particularly, for the passive waveguide components a ratio of a bandwidth density to the corresponding power dissipation was chosen as an ultimate practical characteristic benchmarking their performance. Such a FOM was further expressed in terms of standard 'local' waveguide characteristics, such as signal propagation lengths along straight and curved waveguide sections, cross-talk distance and an optimal bend radius. Interestingly, it provided a general validation of the 'local' FOMs but making different accents on which particular parameters characterising waveguide performance are important (e.g., transmission coefficient vs. propagation length along a curved section). The obtained figure of merit provided an insightful comparison of the performance of the main types of nanophotonic waveguides and shown the ability to efficiently benchmark circuitries on the basis of waveguides with hugely diverse range of characteristics as well as to optimise particular designs. Furthermore, the advantages of the plasmonic approach for the development of plasmonic-based electro-optic modulators have been clearly identified, which establish the communication between electronic and optical domains. The key features of plasmonic-based modulators have been elucidated, namely, their extremely small size which leads to the simultaneous improvement of two key modulator characteristics: the increase of the operational speed and the decrease of the energy consumption per bit. This was demonstrated through derivation of a figure of merit for optical modulators and benchmarking various photonic and plasmonic designs and approaches. Thus, from the prospective of both passive and active functionalities, the plasmonic approach paves the way for the development of highly-efficient hybrid electronic/photonic chips where the information will be processed electronically, but transferred optically. Even now, it can essentially increase the bandwidth of current data communication, despite the increased insertion and propagation losses that are counteracted by the system-level improvement in terms of energy and performance. Further developments, in finding new active materials and optimising the effects for improved electro-optical, thermo-optical and all-optical modulation will benefit both plasmonic and conventional photonic modulators.

**Acknowledgement**
This work was supported by EPSRC (UK), Royal Society, the Wolfson Foundation and US Army Research Office (W911NF-12-1-0533). All data supporting this research are provided in full in the main text of the manuscript.